\documentclass[a4paper,onecolumn,english,epj,nopacs]{svjour}
\usepackage[T1]{fontenc}
\usepackage[latin1]{inputenc}
\usepackage{amsmath}
\usepackage{graphicx}
\usepackage{amssymb}

\makeatletter

\providecommand{\tabularnewline}{\\}
\newcommand{\lyxdot}{.}

\usepackage{color}
\usepackage{bbm}
\usepackage{cite}
\newcommand{\slam}[1]{\setbox0=\hbox{$#1$}
   \dimen0=\wd0
   \setbox1=\hbox{$/$}
   \dimen1=\wd1
   \ifdim\dimen0>\dimen1\hbox{
      \rlap{\hbox to \dimen0{\hfil\box1\hfil}}
      \box0}
   \else\hbox{
      \rlap{\hbox to \dimen1{\hfil\box0\hfil}}
      \box1}
   \fi}
\usepackage{psfrag}

\usepackage{babel}
\makeatother
\begin{document}
\abstract{The Indirect Detection of neutralino Dark Matter is most promising through annihilation channels producing a hard energy spectrum for the detected particles, such as the neutralino annihilation into $Zh$. A cancellation however makes this particular annihilation channel generically subdominant in the huge parameter space of supersymmetric models. This cancellation requires non-trivial relations between neutralino mixings and masses, which we derive from gauge independence and unitarity of the MSSM. To show how the cancellation overshoots leaving only a subdominant result, we use a perturbative expansion in powers of the electroweak/supersymmetry breaking ratio $m_{Z}/m_{\chi}$.}

\title{The Suppression of Neutralino Annihilation into $Zh$}

\author{Labonne Benjamin$^{\textrm{1,2}}$, Nezri Emmanuel$^{\textrm{3}}$,
Orloff Jean$^{\textrm{2}}$}

\institute{$^{\textrm{1}}$Centre de Physique Théorique, CNRS Luminy, case 907,
F-13288 Marseille Cedex 9, France\\
$^{\textrm{2}}$Laboratoire de Physique Corpusculaire, IN2P3-CNRS,
Université Blaise Pascal, F-63177 Aubière Cedex, France\\
$^{\textrm{3}}$Service de Physique Théorique, Université Libre de
Bruxelles, Campus de la Plaine CP225, Belgium}

\mail{labonne@clermont.in2p3.fr, nezri@in2p3.fr, orloff@in2p3.fr}

\date{February, the first, 2006}

\maketitle

\section{Introduction and motivations}

There is no doubt that standard models of particle physics and cosmology
alone cannot describe the full wealth of observational data recently
collected on a wide variety of length scales. Ad-hoc as it may seem,
the Dark Matter (DM) hypothesis \cite{Bertone,Kamionkowski} is probably
part of the minimal set of extra ingredients needed to account for
the increased gravitational self-attraction of matter on scales ranging
from galaxies to the full visible universe. More exotic ingredients
like repulsive Dark Energy, or modifications of gravity itself, might
also become necessary to cope with the apparent acceleration of the
universe. In the absence of a convincing unified theoretical solution
to these both issues, experimental searches are the only way to prove
the validity of hypotheses like the existence of a DM particle. For
instance, the first issue would be settled if a new particle were
found and its non-gravitational interactions measured to be compatible
with the Cold Dark Matter relic density required by Cosmic Microwave
Background and large scale structure formation \cite{tegmark,WMAP}.
However, these interactions are then by definition weak, and we should
not be surprised that their evidences are extremely hard to obtain,
much like for Pauli's neutrino. Debates like the one around DAMA's
claim \cite{DAMA,CDMS-I,CDMS-II,EDELWEIS,EDELWEISFINAL,CRESST-I,CRESST-II,DAMAdiscu}
for direct detection of DM are illustrative of this difficulty. This
is why it is crucial to be able to cross-check and understand results
in as many different ways as possible, for which a definite and well
motivated DM theoretical framework is necessary. In this work, we
shall keep with the well-studied supersymmetric lightest neutralino
($\chi)$ of mSugra or MSSM models.

A particularly crucial cross-check would be the Indirect Detection
of the neutralino annihilation products, which would show that such
annihilation did indeed occur in the past, froze out at some point
and re-started in hot spots like the galactic center or the solar
core, where dark matter later accumulated. However, to identify an
Indirect DM signal and possibly determine the neutralino mass by looking
at fluxes of e.g. photons or neutrinos from such hot spots, it is
essential to be able to distinguish that signal from the standard
but poorly known astrophysical background. These being characterized
by energy spectra with fairly universal power laws, Indirect Detection
will be most successful when neutralino annihilation proceeds through
primary channels which provide secondary photons or neutrinos with
the hardest possible spectra, and a sharp energy cut-off around the
neutralino mass. From fig.~\ref{nugammaflux} (discussed in Appendix
\ref{sec:Indirect-Detection-Energy}), the most promising annihilation
channels are into a $\tau^{+}\tau^{-}$ pair, into two gauge bosons
($\chi\chi\to W^{+}W^{-}$, or $ZZ$ which has the same shape) or
into one gauge boson and a Higgs-Englert-Brout (HEB) boson\cite{Brout-Englert,Higgs}
($\chi\chi\to Zh$).%
\begin{figure}
\begin{centering}\psfrag{Egammamx}[rt][rB]{$x=E_\nu/m_\chi$}

\psfrag{dNdx}[rB][rt]{$dN/dx$}

\psfrag{mass}[c]{$m_{\chi}=1$ TeV}

\psfrag{process}{}

\psfrag{10}{10}

\psfrag{1}{1}

\psfrag{-1}{\tiny -1}

\psfrag{-2}{\tiny -2}

\psfrag{2}{\tiny 2}

\psfrag{3}{\tiny 3}

\psfrag{Zh}{\color[rgb]{1,0,1}$Zh$}

\psfrag{W+W-}{$W^+W^-$}

\psfrag{bbbar}{\color[rgb]{1,0,0}$b\bar b$}

\psfrag{ttbar}{\color[rgb]{0,0.7,0}$t\bar t$}

\psfrag{tau+tau-}{\color[rgb]{0,0,1}$\tau^+\tau^-$}\includegraphics[width=0.95\columnwidth,height=0.55\columnwidth]{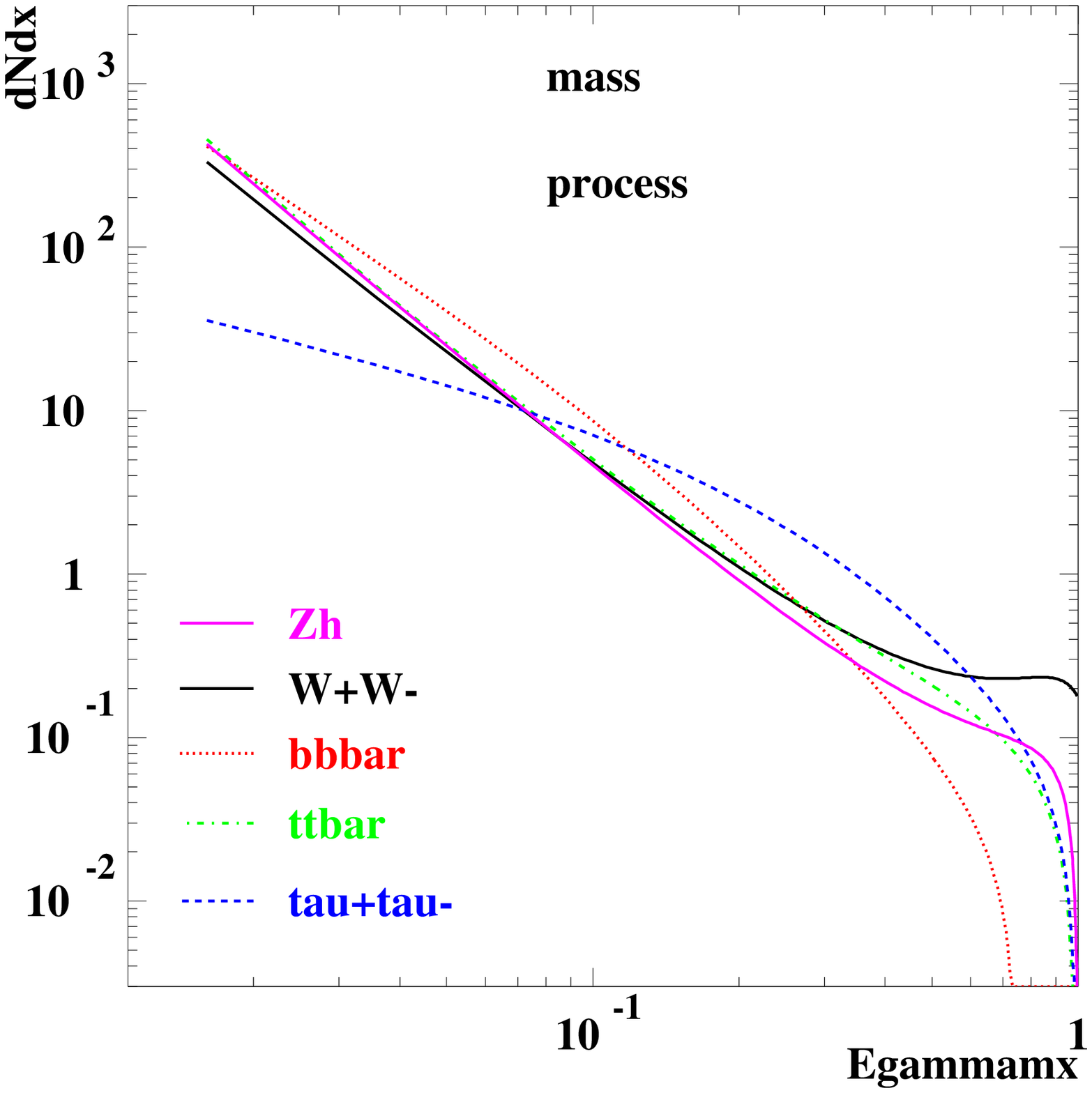}\psfrag{Egammamx}[rt][rB]{$x=E_\gamma/m_\chi$}

\psfrag{dNdx}[rB][rt]{$dN/dx$}

\psfrag{mass}[c]{$m_{\chi}=1$ TeV}

\psfrag{process}{}

\psfrag{10}{10}

\psfrag{1}{1}

\psfrag{-1}{\tiny -1}

\psfrag{-2}{\tiny -2}

\psfrag{2}{\tiny 2}

\psfrag{3}{\tiny 3}

\psfrag{Zh}{\color[rgb]{1,0,1}$Zh$}

\psfrag{W+W-}{$W^+W^-$}

\psfrag{bbbar}{\color[rgb]{1,0,0}$b\bar b$}

\psfrag{ttbar}{\color[rgb]{0,0.7,0}$t\bar t$}

\psfrag{tau+tau-}{\color[rgb]{0,0,1}$\tau^+\tau^-$}\includegraphics[width=0.95\columnwidth,height=0.55\columnwidth]{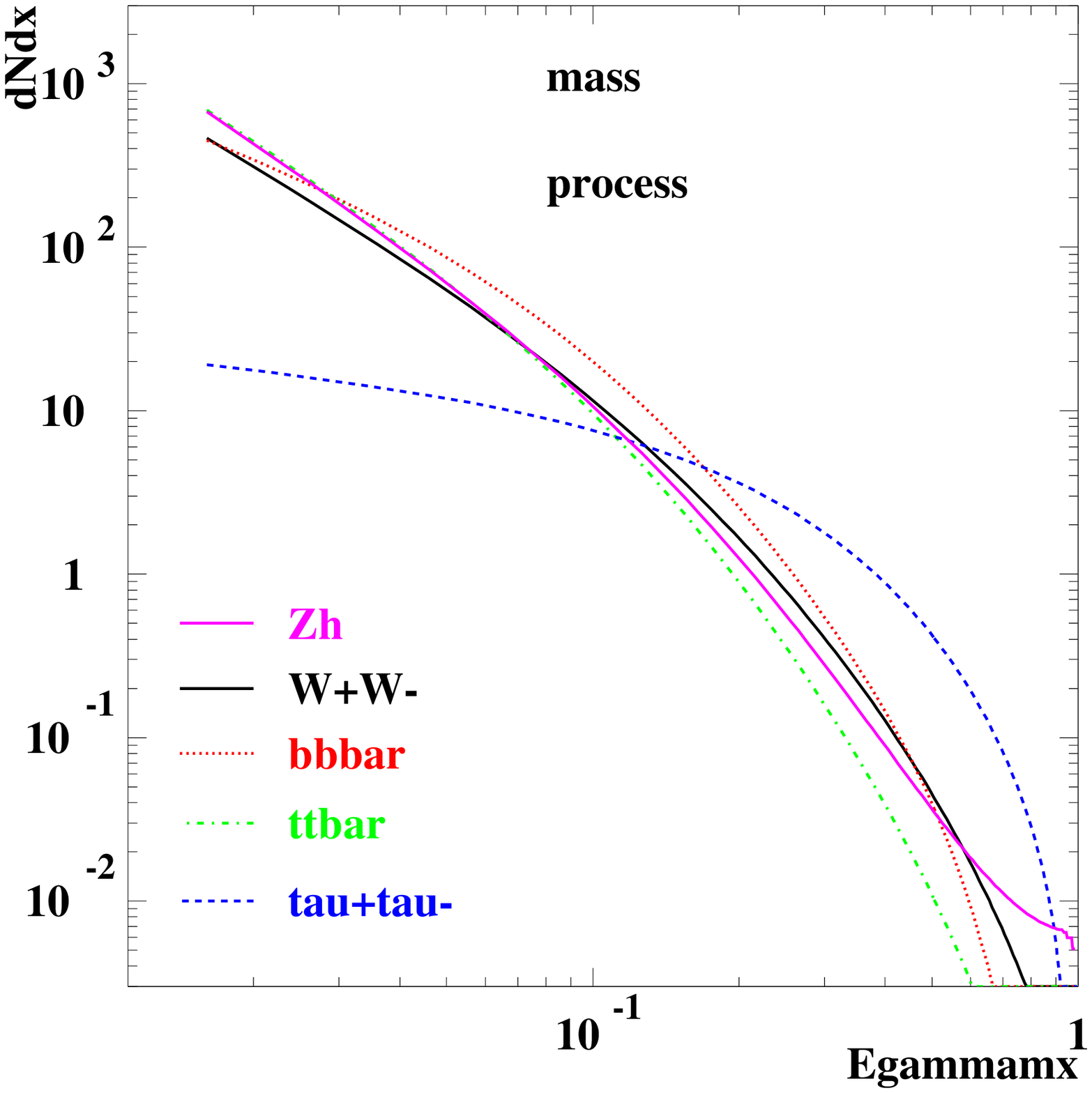}\par\end{centering}

\caption{The shape of differential neutrino (top) and gamma (bottom) fluxes
coming from neutralino annihilations into various primary channels.
The normalisations (depending on branching ratios) have been arbitrarily
rescaled to compare the shape of all channels; the dependence on $m_{\chi}$
($1\ \textrm{TeV}$ here) is weak.\label{nugammaflux}}
\end{figure}

\begin{figure}
\begin{centering}\psfrag{B}{\hspace{-3mm} $m_{0}$ (\textrm{GeV})}

\psfrag{M}{\vspace{-0.5cm} $\begin{array}{c} m_{1/2} \\ (\textrm{GeV}) \end{array}$}

\psfrag{C}{$\frac{\sigma\left(\chi\chi\rightarrow Zh\right)_{Z+\chi}}{\sigma\left(\chi\chi\rightarrow All\right)}$}\includegraphics[width=0.9\columnwidth,keepaspectratio]{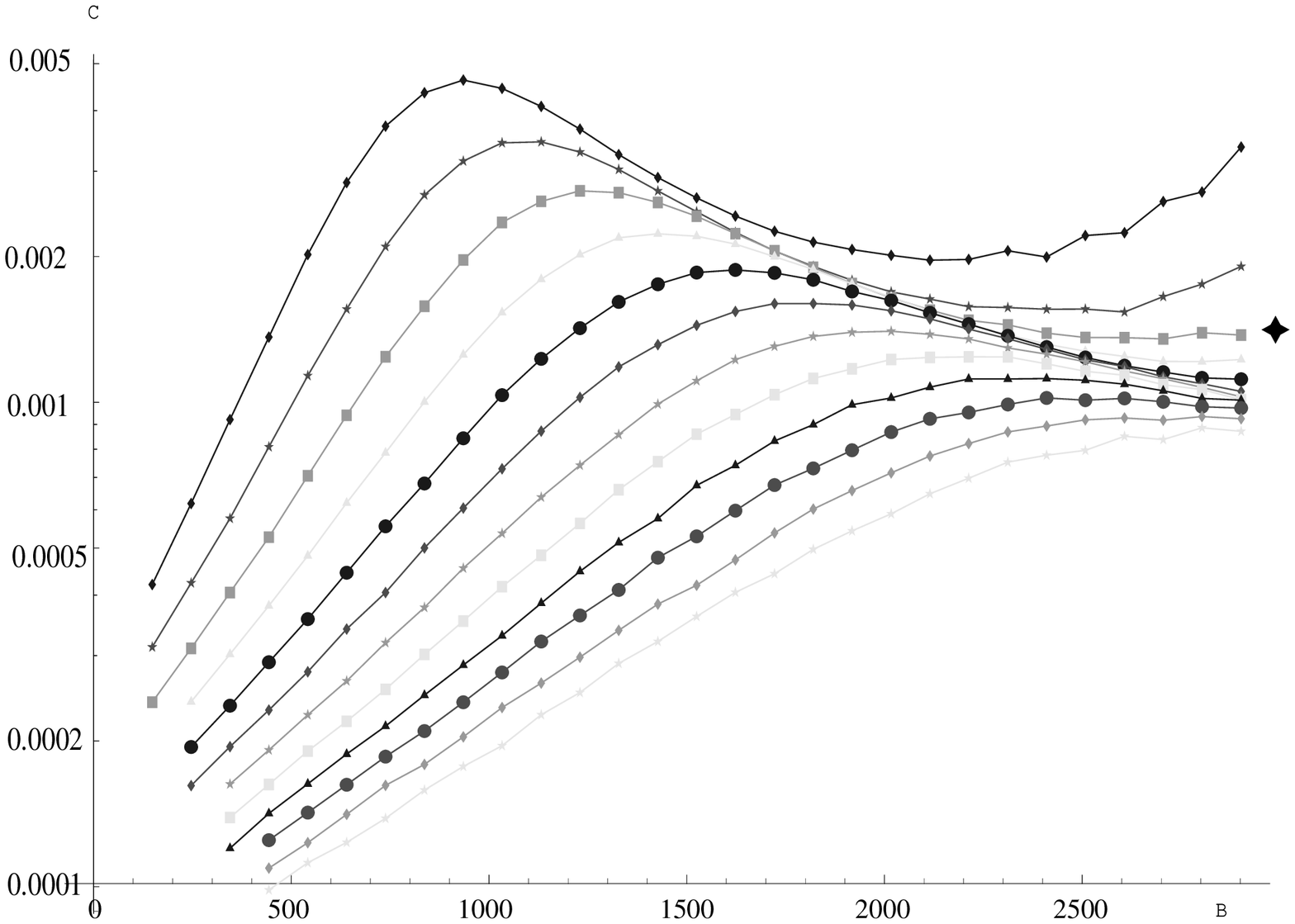}\par\end{centering}

\vspace{0,5cm}

\psfrag{A}{$\frac{\sigma\left(\chi\chi\rightarrow Zh\right)_Z}{\sigma\left(\chi\chi\rightarrow All\right)}$}

\psfrag{B}{$m_{0}$ (\textrm GeV)}

\psfrag{M}{\vspace{-0.5cm} $\begin{array}{c} m_{1/2} \\ (\textrm GeV) \end{array}$}

\begin{centering}\includegraphics[width=0.85\columnwidth,keepaspectratio]{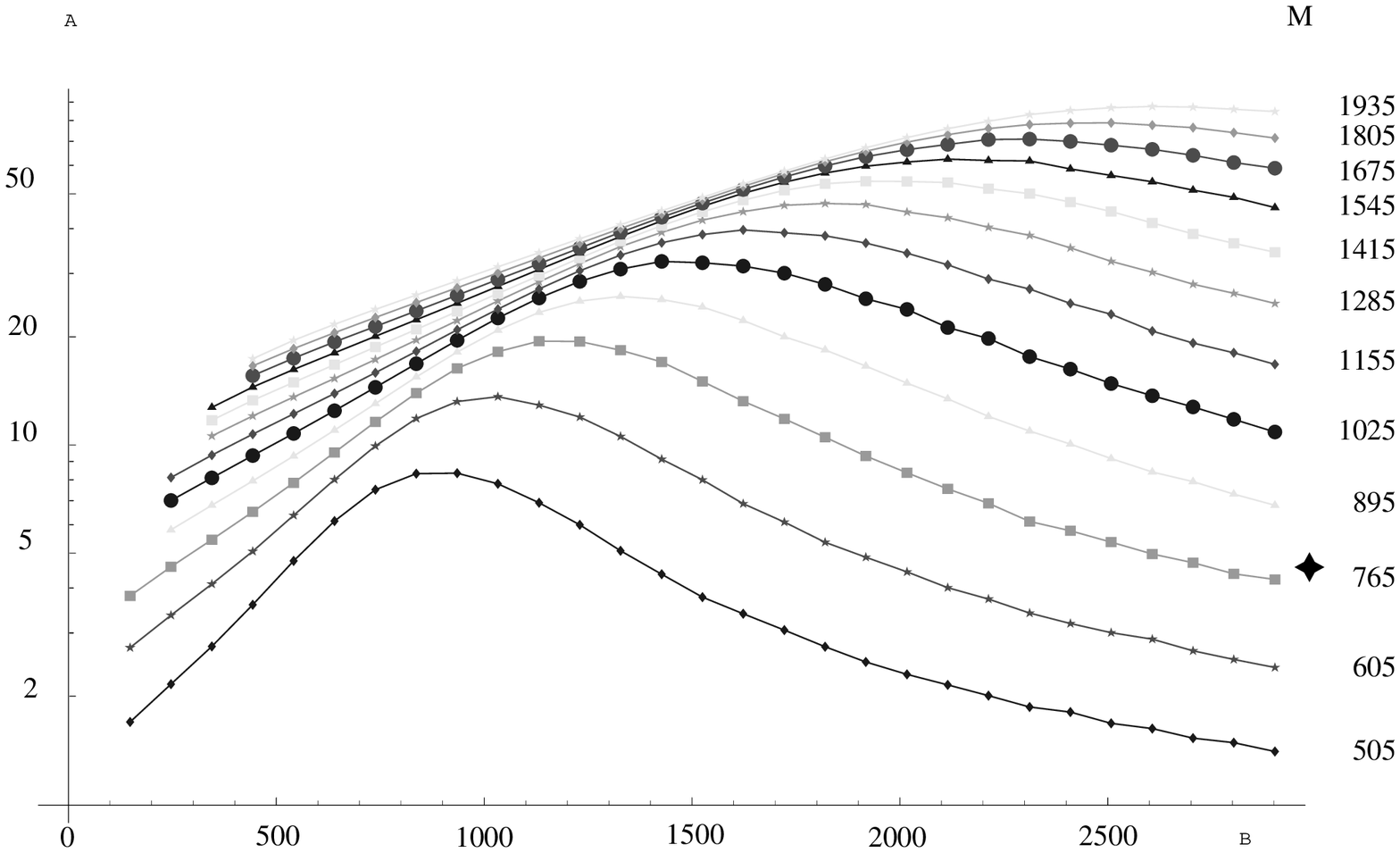}\par\end{centering}

\caption{Neutralino annihilation branching ratios to $Zh$ for various $m_{0}$
and $m_{1/2}$ values: $Z$-exchange only (bottom), $Z$ and $\chi$
exchanges (top)\label{Zsursigamtot}}
\end{figure}

However, these fairly universal spectra need to be weigh\-ted by
the actual model-dependent branching ratios to give the final indirect
DM detection signal. It was noted long ago\cite{Drees} that the $Zh$
channel is then suppressed, which can be numerically checked%
\footnote{Temporarily failing such check after an update of the DarkSusy-Suspect
interface was actually the starting point of this work.%
} using the DarkSusy (3.14.02 version)\cite{Darksusy} and the Suspect
(2.003 version) code\cite{Suspect}: the top plot of fig.~\ref{Zsursigamtot}
typically shows a suppression by three orders of magnitude for various
mSugra models with $\tan\beta=10$. To qualify this suppression, let
us start by contributions we expect to be dominant, namely those from
SM particle exchanges (in our case an $s$-channel $Z$) since superpartners
are necessarily heavier. As seen in the bottom fig.~\ref{Zsursigamtot},
this contribution not only dominates, but even overwhelms the total
cross-section. We therefore need to understand a double suppression,
that first cancels this large contribution and second brings the total
$Zh$ annihilation below other channels. The cancelling contribution
necessarily involves non-SM particle exchanges, which seems contradictory
with the fact that on fig.~\ref{Zsursigamtot}, the cancellation
gets better with increasing $m_{0}$ and $m_{1/2}$, \emph{i.e.} for
maximally broken supersymmetry. %

\begin{figure}
\begin{centering}\psfrag{C}{}

\psfrag{A}{$\mu$ (\textrm GeV)}

\psfrag{B}{$\begin{array}{c} M_{1} \\ (\textrm GeV) \end{array}$}

\psfrag{E}{}

\psfrag{D}{}

\psfrag{-5}{$10^{-5}$}

\psfrag{-4}{$10^{-4}$}

\psfrag{-3}{$10^{-3}$}

\psfrag{-2}{$10^{-2}$}

\psfrag{-1}{$10^{-1}$}

\psfrag{0}{$1$}\includegraphics[width=0.95\columnwidth,keepaspectratio]{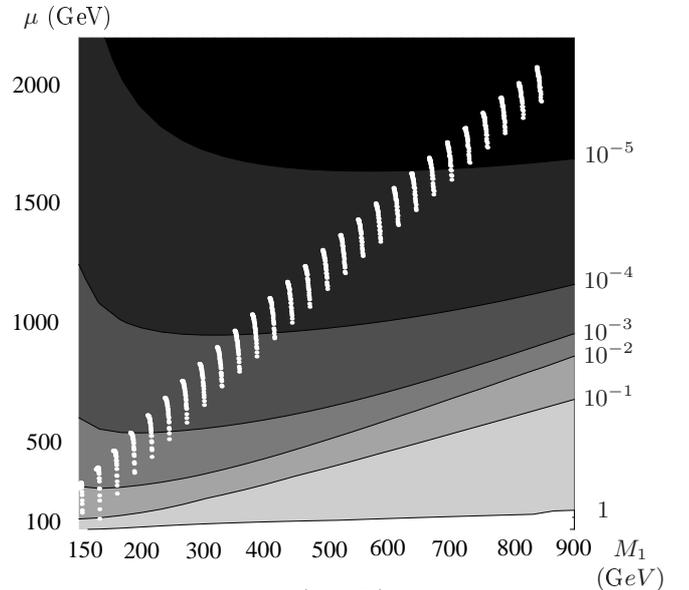}\vspace{2mm}\par\end{centering}

\caption{Contour plot of $\frac{\sigma\left(\chi\chi\rightarrow Zh\right)_{Z+\chi}}{\sigma\left(\chi\chi\rightarrow Zh\right)_{Z}}$
in the MSSM; white dots correspond to mSugra models of fig.~\ref{Zsursigamtot}.\label{MSSMscan}}
\end{figure}

To be as general as possible, this cancellation should be checked
in a supersymmetry breaking independent way. This is done in fig.~\ref{MSSMscan}
in the more general MSSM, while keeping the usual GUT relation $M_{1}=\frac{5}{3}\tan\theta_{W}M_{2}\simeq0.5M_{2}$.
A cancellation up to three orders of magnitude thus seems a generic
property of every broken supersymmetry theory. %

To be more concrete, let us now focus on the particular mSugra model
with $m_{0}=3000\ GeV$, $m_{1/2}=800\ GeV$, $A_{0}=0$, $\tan\left(\beta\right)=10$
and $\mu>0$, marked by the black star in fig.~\ref{Zsursigamtot}.
The annihilation cross-sections at rest are:

\begin{center}\begin{tabular}{|c|c|c|}
\hline 
$v\sigma\ \left(cm^{3}/s\right)$&
$\chi\chi\rightarrow t\overline{t}$&
$\chi\chi\rightarrow Zh$\tabularnewline
\hline
$Z$ exchange &
$1.83\times10^{-28}$&
$4.72\times10^{-28}$\tabularnewline
\hline 
All diagrams&
$1.03\times10^{-28}$&
$1.48\times10^{-31}$\tabularnewline
\hline 
\multicolumn{3}{|c|}{$v\sigma\left(\chi\chi\rightarrow\mathrm{all}\right)=1.05\times10^{-28}$}\tabularnewline
\hline
\end{tabular}\par\end{center}

The annihilation is dominated by the $t\bar{t}$ channel, 3 orders
of magnitude larger than the $Zh$ one \cite{Nezri}. However, when
restricting to the $Z$ exchange diagram, they are comparable, with
$Zh$ slightly larger. This is easily understood by exhibiting the
couplings and kinematic factors in the amplitudes:

\begin{center}\begin{tabular}{cc}
\includegraphics{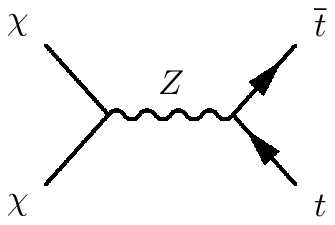}\vspace{0,3cm}&
\includegraphics{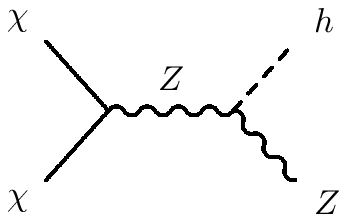}\vspace{0,3cm}\tabularnewline
$A\left(\chi\chi\rightarrow t\overline{t}\right)_{Z}\propto\frac{-g^{2}m_{t}O_{11}^{Z}}{\cos^{2}\theta_{W}}$&
$A\left(\chi\chi\rightarrow Zh\right)_{Z}\propto\frac{-g^{2}m_{Z}O_{11}^{Z}}{\cos^{2}\theta_{W}}$\tabularnewline
\end{tabular}\par\end{center}

\noindent where the $\chi$$\chi$$Z$ coupling $O_{11}^{Z}$ is defined
in terms of the neutralino mixing matrix%
\footnote{In what follow, we will always work with the hypothesis of CP violating
phases absence, in order the $N$ matrix to be real (see Appendix
\ref{sec:Neutralino-Mass-Matrix})%
} $N$ (see Appendix \ref{sub:Z-chi-chi-coupling}):

\begin{eqnarray}
O_{ij}^{Z} & = & \left(N_{i4}N_{j4}-N_{i3}N_{j3}\right)\label{OZij}\end{eqnarray}

It remains to understand why other exchanges cancel the $Zh$ channel
and not the $t\overline{t}$ one. For the latter, the $t$-channel
sfermion exchange can be made arbitrarily small by taking large enough
$m_{0}$, so that the main contribution proceeds via a (SM) $Z$ boson
exchange, conforming to naive expectations. However, two other diagram
are involved in the $Zh$ annihilation channel: (1) an $s$-channel
pseudoscalar $A$ exchange, which can be neglected for large $m_{0}$
(in what follows we will always work in this pseudoscalar decoupling
limit (\ref{decoupling-condition})), and (2) $t$-channel exchanges
of the 4 neutralinos, which cannot be decoupled because of strong
links between the couplings involved in each diagram. These links
appear in the expression of the annihilation amplitudes\cite{jungman,Roszkowski}
derived in the next section: 

\begin{eqnarray}
A\left(\chi\chi\rightarrow Zh\right)_{Z} & = & \frac{-ig^{2}\sqrt{2}}{\cos^{2}\theta_{W}}\frac{m_{\chi}^{2}}{m_{Z}^{2}}\beta_{Zh}\times O_{11}^{Z}\label{amplZ}\\
A\left(\chi\chi\rightarrow Zh\right)_{\chi} & = & \frac{ig^{2}\sqrt{2}}{\cos^{2}\theta_{W}}\frac{m_{\chi}^{2}}{m_{Z}^{2}}\beta_{Zh}\times\label{amplchi}\\
 &  & \times\sum_{i=1}^{4}\frac{2O_{1i}^{Z}O_{1i}^{h}\left(m_{\chi_{i}}-m_{\chi}\right)m_{Z}}{2m_{\chi}^{2}+2m_{\chi_{i}}^{2}-m_{h}^{2}-m_{Z}^{2}}\nonumber \end{eqnarray}
where the $\chi_{i}$$\chi_{j}$$h$ coupling $O_{ij}^{h}$ is defined
in Appendix (\ref{Ch}) as (where the decoupling condition (\ref{decoupling-condition})
has been used):\begin{eqnarray}
\hspace{-2mm}O_{ij}^{h} & = & \left(c_{W}N_{i2}-s_{W}N_{i1}\right)\left(s_{\beta}N_{j4}-c_{\beta}N_{j3}\right)+\left(i\leftrightarrow j\right)\label{Ohij}\end{eqnarray}
It is clear that the second amplitude (\ref{amplchi}) cannot easily
be neglected and might turn out comparable with the first one (\ref{amplZ}).
However, it is less clear why both should cancel with high precision,
especially given as different-looking couplings as (\ref{OZij}) and
(\ref{Ohij}).

A toy example of the above cancellation is provided by the annihilation
of a spin singlet $t\bar{t}$ pair into $Zh$ in a Standard Model
without $SU(2)$. One may first be puzzled to get a cancellation between
an $s$-channel $Z$-exchange, which only involves gauge couplings,
and a $t$-channel top exchange, which involves an a priori independent
Yukawa coupling. However, one soon realizes that the existence of
the $s$-channel requires both spontaneous breaking of the gauge symmetry
for the $ZZh$ vertex, and an axial coupling for the $t\bar{t}Z$
vertex. This last coupling excludes contributions to the top mass
other than the gauge breaking $y\langle h\rangle$ one. The two channels
$g\frac{1}{m_{Z}^{2}}gm_{z}\propto g\frac{1}{m_{t}}y$ are then both
proportional to $g/\langle h\rangle$ and may cancel. In contrast
with this simple case, the neutralino annihilation studied below is
complicated by the presence of another source of mass, namely the
SUSY breaking Majorana mass terms for the gauginos.

To analyse this cancellation, we start in section 2 by deriving the
relevant amplitudes. In section 3, gauge independence is used to draw
a first link between the couplings, which is shown to follow from
the gauge invariance of the mass matrix. Unitarity at high energy
is then used in section 4 to derive a second relation, which is combined
with the first one to show that a cancellation is possible. In section
5, the structure of Rayleigh-Schroedinger perturbation theory is then
used to show that this cancellation is stronger than expected.

\section{Amplitudes for Neutralino Annihilation at Rest}

Let us start by deriving the polarized amplitude $A\left(Zh\right)_{Z}$
for neutralino annihilation into $Zh$ via $Z$ exchange. By construction,
neutralinos are their own antiparticles, so that an initial pair of
lightest neutralinos at rest is necessarily in an antisymmetric spin
singlet state. The final state containing a HEB scalar, the outgoing
$Z$ boson polarization needs to be longitudinal, and can be chosen
as $z$-axis. It is then more convenient to use helicity amplitudes\cite{Drees,jungman}
than unpolarized cross-sections\cite{Roszkowski}. The Feynman diagram
and rules defined in the Appendix \ref{sec:Lagrangian-terms-for}
give the amplitude

\begin{center}\includegraphics{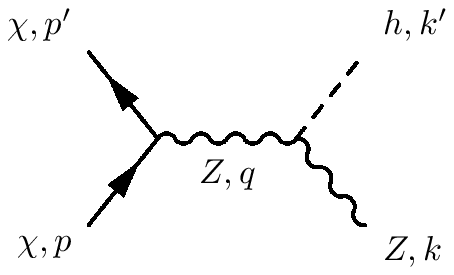}\par\end{center}

\vspace{-6mm}

\begin{eqnarray}
A\left(\chi\overline{\chi}\rightarrow Zh\right)_{Z} & = & \frac{-iC_{11}^{A}C_{Z}}{q^{2}-m_{Z}^{2}}\left(\overline{\chi}\left(p'\right)\gamma^{\mu}\gamma_{5}\chi\left(p\right)\right)\label{firstamplitudeZ}\\
 &  & \hphantom{\frac{-iC_{11}^{A}C_{Z}}{q^{2}-m_{Z}^{2}}}\times\left(g_{\mu\nu}-\frac{q_{\mu}q_{\nu}}{m_{Z}^{2}}\right)\varepsilon^{\nu}\left(k\right)\nonumber \end{eqnarray}
where $\chi=u$ is the incoming neutralino, $\overline{\chi}=\overline{v}$
the outgoing one in an arbitrary choice of arrows directions. $u$
and $v$ are external Dirac spinors at rest in the chiral basis (see
Appendix \ref{sub:Dirac-Majorana}), namely:\begin{eqnarray*}
\chi\left(p_{0}\right) & = & \sqrt{m_{\chi}}\left(\begin{array}{c}
\xi_{s}\\
\xi_{s}\end{array}\right);\mathrm{\ with\ }\xi_{+\frac{1}{2}}=\left(\begin{array}{c}
-1\\
0\end{array}\right)\mathrm{\ or\ }\xi_{-\frac{1}{2}}=\left(\begin{array}{c}
0\\
1\end{array}\right)\end{eqnarray*}
for a spin up or down along the $z$-axis, and similarly for the {}``antiparticle'':

\begin{eqnarray*}
\overline{\chi}\left(p_{0}'\right) & = & \sqrt{m_{\chi}}\left(-\eta_{s'},\eta_{s'}\right)\end{eqnarray*}
Thanks to the Majorana condition (\ref{majorana-v}), descriptions
of the same neutralino as a particle with $\xi_{s'}$ or as an {}``antiparticle''
with $\eta_{s'}$ are equivalent provided:\begin{eqnarray}
\eta_{s'} & = & -i\sigma^{2}\xi_{s'}\label{majorana-condition-chiral}\end{eqnarray}
The polarization of the longitudinal $Z$ boson being $\varepsilon^{\nu}(k)=(k^{z},0,0,k^{0})/m_{Z}$,
and the initial momentum at rest simply $q^{\mu}=(2m_{\chi},\vec{0})$,
the polarized amplitude (\ref{firstamplitudeZ}) is:

\begin{eqnarray}
A\left(\chi_{\uparrow}\chi_{\downarrow}\rightarrow Zh\right)_{Z} & = & \frac{iC_{11}^{A}C_{Z}}{4m_{\chi}^{2}-m_{Z}^{2}}\frac{m_{\chi}}{m_{Z}}(\sigma_{11}^{\mu}+\bar{\sigma}_{11}^{\mu})\label{spinup}\\
 &  & \times\left(g_{\mu\nu}-\frac{q_{\mu}q_{\nu}}{m_{Z}^{2}}\right)\left(k^{z},0,0,k^{0}\right)\nonumber \\
 & = & -2iC_{11}^{A}C_{Z}\frac{m_{\chi}}{m_{Z}^{3}}k^{z}\label{ampk}\end{eqnarray}

Notice the time-like structure of the initial state vector $(\sigma_{11}^{\mu}+\bar{\sigma}_{11}^{\mu})=(2,0,0,0)$:
a purely axial coupling talks only with the scalar part of the two
spins at rest. Notice also the disappearing of the $Z$ pole at $m_{\chi}=m_{Z}/2$.

Expressing $k^{z}=m_{\chi}\beta_{Zh}$ in terms of the conventional
kinematic factor

\[
\beta_{Zh}=\sqrt{1-\frac{\left(m_{h}+m_{Z}\right)^{2}}{4m_{\chi}^{2}}}\sqrt{1-\frac{\left(m_{h}-m_{Z}\right)^{2}}{4m_{\chi}^{2}}}\begin{array}[t]{c}
\approx\\
m_{\chi}\gg m_{Z}\end{array}1\]
and using the definitions (\ref{Cv}) and (\ref{Cz}) of couplings
in terms of neutralino mixings given in the Appendix \ref{sec:Lagrangian-terms-for},
we finally find:

\begin{eqnarray}
A\left(\chi_{\uparrow}\chi_{\downarrow}\rightarrow Zh\right)_{Z} & = & \frac{-ig^{2}O_{11}^{Z}}{\cos^{2}\theta_{W}}\frac{m_{\chi}^{2}}{m_{Z}^{2}}\beta_{Zh}\label{spinupfin}\end{eqnarray}

To get the amplitude with reversed helicities, we just need to replace
(1,1) components of the Pauli matrices in (\ref{spinup}) by (2,2)
components, and take the extra sign from the Majorana condition (\ref{majorana-condition-chiral})
into account. Thanks to this sign, only an antisymmetric initial state
can contribute, and gives with the correct normalization factor:

\begin{eqnarray}
A\left(\chi\chi\rightarrow Zh\right)_{Z} & = & \frac{-ig^{2}\sqrt{2}}{\cos^{2}\theta_{W}}\frac{m_{\chi}^{2}}{m_{Z}^{2}}\beta_{Zh}O_{11}^{Z}\label{ampZhnew}\end{eqnarray}
in agreement with (\ref{amplZ}) and existing results\cite{jungman,Roszkowski}. 

This amplitude (\ref{ampZhnew}) has to be compared with the result
of a similar computation for the annihilation into $t\bar{t}$:

\begin{eqnarray}
\hspace{-0.5cm}A\left(\chi\chi\rightarrow t\overline{t}\right)_{Z} & = & \frac{-ig^{2}\sqrt{2}}{\cos^{2}\theta_{W}}\frac{m_{\chi}m_{t}}{m_{Z}^{2}}\beta_{t\overline{t}}T_{3}O_{11}^{Z}\label{ampttbarnew}\end{eqnarray}
where $\beta_{t\bar{t}}=\sqrt{1-m_{t}^{2}/m_{\chi}^{2}}$ and $T_{3}$=1
is the weak isospin of the top quark. Notice the different power of
$m_{\chi}$, favouring the $Zh$ channel for large masses.

We have seen in the introduction that $t$-channel neutralinos exchanges
should reverse this conclusion. Following the same path, their contribution
is 

\begin{center}\includegraphics{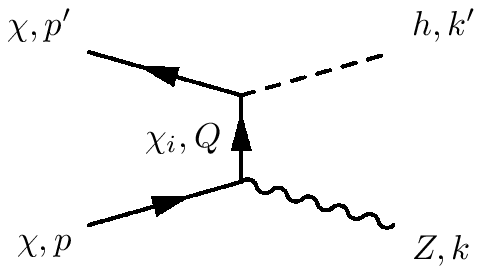}\vspace{-0.6cm}\par\end{center}

\begin{eqnarray*}
A\left(\chi\chi\rightarrow Zh\right)_{\chi_{i}} & = & \frac{-iC_{1i}^{A}C_{1i}^{h}}{Q^{2}-m_{\chi_{i}}^{2}}\\
 &  & \times\left(\overline{\chi}\left(p'\right)\left(\slam{Q}+m_{\chi_{i}}\right)\gamma^{\mu}\gamma_{5}\chi\left(p\right)\right)\varepsilon_{\mu}\left(k\right)\end{eqnarray*}
with $Q^{\mu}=p^{\mu}-k^{\mu}$. After some algebra on the $t$-channel,
and adding the $u$-contribution (same diagram as before with $p$
and $p'$ interchanged), one finds:

\begin{eqnarray}
A\left(\chi\chi\rightarrow Zh\right)_{\chi_{i}} & = & \frac{ig^{2}\sqrt{2}}{\cos^{2}\theta_{W}}\frac{m_{\chi}^{2}}{m_{Z}^{2}}\beta_{Zh}\times\label{amplitudecchii}\\
 &  & \times\frac{2O_{1i}^{Z}O_{1i}^{h}\left(m_{\chi_{i}}-m_{\chi}\right)m_{Z}}{2m_{\chi}^{2}+2m_{\chi_{i}}^{2}-m_{Z}^{2}-m_{h}^{2}}\nonumber \end{eqnarray}
 The problem is now to understand how the amplitudes (\ref{amplitudecchii})
and (\ref{ampZhnew}) cancel with the $10^{-3}$ precision shown in
the introduction. This can happen only if the sum of second lines
of (\ref{amplitudecchii}) which contain four powers of the neutralino
mixing matrix $N$ somehow reduces to $O_{11}^{Z}\sim N^{2}$ as a
consequence of some symmetry. We already noticed that supersymmetry
had to be maximally broken for the cancellation to take place. We
are thus left with gauge invariance which is investigated in the next
section.

\section{Gauge Independence and Gauge Invariance of the Mass Matrix}

The previous computations were performed in the unitary gauge, where
massive gauge fields have completely {}``eaten'' a Goldstone boson.
One way to obtain non-trivial relations of the kind we seek is to
work in the $R_{\xi}$-gauge family of 't Hooft\cite{Rchi}, and require
independence of the result on the gauge-fixing parameter $\xi$.

Let us first notice that the neutralino exchange diagrams are gauge
independent by themselves. Indeed, in a vector supermultiplet, only
the bosonic gauge field is gauge dependent, and not the associated
gaugino. Moreover, higgsinos are associated with the real part of
complex scalars, which are also gauge independent.

We can thus concentrate on the $Z$ exchange diagram. When going from
unitary to $R_{\xi}$-gauges, this contribution splits in two $\xi$-dependent
diagrams: one with Goldstone boson exchange, and another with the
$Z$-boson exchange. To exhibit the cancellation of their $\xi$ dependence,
it is convenient in the $Z$ propagator 

\begin{eqnarray*}
\frac{-i}{q^{2}-m_{Z}^{2}}\left(g_{\mu\nu}-\frac{\left(1-\xi\right)q_{\mu}q_{\nu}}{q^{2}-\xi m_{Z}^{2}}\right)\end{eqnarray*}
to decompose the longitudinal part as:\begin{eqnarray}
\hspace{-0.5cm}\frac{m_{Z}^{2}\left(1-\xi\right)}{\left(q^{2}-m_{Z}^{2}\right)\left(q^{2}-\xi m_{Z}^{2}\right)} & = & \frac{1}{\left(q^{2}-m_{Z}^{2}\right)}-\frac{1}{\left(q^{2}-\xi m_{Z}^{2}\right)}.\label{propagatorRchi}\end{eqnarray}
The first term is nothing but the longitudinal unitary gauge propagator,
and the second exhibits a fake pole at the Goldstone mass $q^{2}=\xi m_{Z}^{2}$,
with a wrong sign \cite{cornwall}. The $\xi$-dependent $Z$ exchange
$A_{Z\xi}=A_{Z}+A_{ZG}$ correspondingly decomposes into the previously
obtained $A_{Z}$ (\ref{firstamplitudeZ}) and a Goldstone-like amplitude
with gauge couplings:\begin{eqnarray*}
A & \left(\chi\chi\rightarrow Zh\right)_{ZG}= & -i\frac{C^{A}C_{Z}}{m_{Z}^{2}}\left(\overline{\chi}\left(p'\right)\slam{q}\gamma_{5}\chi\left(p\right)\right)\frac{q.\varepsilon\left(k\right)}{q^{2}-\xi m_{Z}^{2}}\end{eqnarray*}
which, using the mass-shell condition for the initial state $\bar{\chi}\slam{q}\gamma_{5}\chi=2m_{\chi}\bar{\chi}\gamma_{5}\chi$,
becomes: 

\begin{eqnarray}
A & \left(\chi\chi\rightarrow Zh\right)_{ZG}= & \frac{-ig^{2}O_{11}^{Z}}{2\cos^{2}\theta_{W}}\frac{m_{\chi}}{m_{Z}}\overline{\chi}\gamma_{5}\chi\frac{q.\varepsilon\left(k\right)}{q^{2}-\xi m_{Z}^{2}}\label{AmpZxi}\end{eqnarray}

We now turn to the genuine Goldstone boson exchange. The $ZZh$ coupling
is replaced by the $GZh$ one: 

\begin{center}\includegraphics{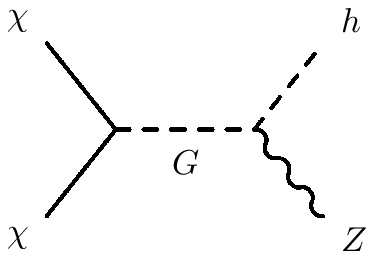}\vspace{-0,6cm}\par\end{center}

\begin{eqnarray}
C_{Gh}^{Z} & = & iC_{G}\left(k-k'\right)_{\mu}where\ C_{G}=g/2\cos\theta_{W}\label{GZhcoupling}\end{eqnarray}
and the scalar propagator is simply $i/(q^{2}-\xi m_{Z}^{2})$. But
things are more subtle for the $G\chi\chi$ vertex: even if the neutral
Goldstone boson is part of the $Z$ boson in the unitary gauge, it
does not have the same coupling to neutralinos:

\begin{eqnarray}
C_{\chi_{i}\chi_{j}}^{G} & = & \frac{igO_{ij}^{G}}{2\cos\theta_{W}}\label{G-neutralino-neutralino-coupling}\end{eqnarray}
with

\begin{equation}
O_{ij}^{G}=\left(N_{i2}c_{W}-N_{i1}s_{W}\right)\left(c_{\beta}N_{j3}+s_{\beta}N_{j4}\right)+{\scriptstyle \left(i\leftrightarrow j\right)}\label{OGij}\end{equation}
For the lightest neutralino annihilation, $i=j=1$, and since the
coupling is imaginary, the Goldstone only couples to the axial part
of the neutralino. The amplitude for the Goldstone exchange diagram
is thus: 

\begin{eqnarray}
A & \left(\chi\chi\rightarrow Zh\right)_{G}= & -C_{\chi\chi}^{G}C_{G}\ \overline{\chi}\left(p'\right)\gamma_{5}\chi\left(p\right)\label{amp-goldstone-xi-gauge}\\
 &  & \hphantom{-C_{\chi\chi}^{G}C_{G}}\times\frac{\left(q+k'\right)_{\mu}}{q^{2}-\xi m_{Z}^{2}}\varepsilon^{\mu}\left(k\right)\nonumber \end{eqnarray}
Recalling kinematics $\left(q=k+k'\right)$, the polarization condition
$\left(k.\varepsilon\left(k\right)=0\right)$, and couplings definitions,
we finally have: 

\begin{eqnarray}
\begin{array}{c}
A\left(\chi\chi\rightarrow Zh\right)_{G}\end{array} & \begin{array}{c}
=\end{array} & \begin{array}{c}
i\frac{g^{2}O_{11}^{G}}{2c_{W}^{2}}\end{array}\ \overline{\chi}\gamma_{5}\chi\ \frac{q.\varepsilon\left(k\right)}{q^{2}-\xi m_{Z}^{2}}\label{AmpGxi}\end{eqnarray}
Now, comparing (\ref{AmpZxi}) and (\ref{AmpGxi}), gauge independence
requires a relation between couplings: 

\begin{eqnarray*}
O_{11}^{Z}\frac{m_{\chi}}{m_{Z}} & = & -\frac{1}{2}O_{11}^{G}\end{eqnarray*}
which by (\ref{OZij}) and (\ref{OGij}) can be expressed in terms
of the neutralino mixings and masses:

\begin{eqnarray}
 &  & \left(N_{14}^{2}-N_{13}^{2}\right)\frac{m_{\chi}}{m_{Z}}\label{rela-gauge}\\
 &  & \hspace{1em}=-\left(N_{12}c_{W}-N_{11}s_{W}\right)\left(s_{\beta}N_{14}+c_{\beta}N_{13}\right)\nonumber \end{eqnarray}
This relation can be extended for the annihilation of an arbitrary
pair of neutralinos $\chi_{i}-\chi_{j}$in the same channel: \begin{eqnarray*}
O_{ij}^{Z}\frac{m_{\chi_{i}}+m_{\chi_{j}}}{m_{Z}} & = & -O_{ij}^{G}\end{eqnarray*}
 which is a short version of the rather non-trivial identity:

\begin{eqnarray}
 &  & \left(N_{i4}N_{j4}-N_{i3}N_{j3}\right)\frac{m_{\chi_{i}}+m_{\chi_{j}}}{m_{Z}}\label{general-gauge-relation}\\
 &  & \hspace{1cm}=-\left(N_{i2}c_{W}-N_{i1}s_{W}\right)\left(s_{\beta}N_{j4}+c_{\beta}N_{j3}\right)-{\scriptstyle \left(i\leftrightarrow j\right)}\nonumber \end{eqnarray}

Although not completely identical, this relation bears similarities
with the combination of masses and mixings involved in (\ref{amplchi}).
It is therefore interesting to notice that it only involves the neutralino
mass matrix, and can be derived in the following way. By definition
of the mixing matrix $N$ (\ref{diagonalization-formula}), we have:\begin{eqnarray*}
\left(NM\right)_{ij} & = & m_{i}N_{ij}\end{eqnarray*}
Then for any matrix $P$, the following identities hold:

\begin{eqnarray}
\hspace{-0.5cm}\left(m_{i}+m_{j}\right)\left(NPN^{-1}\right)_{ij} & = & \left(N\left(MP+PM\right)N^{-1}\right)_{ij}\label{relamatrix}\end{eqnarray}
As a particular case, if we take for $P$ the isospin operator that
flips the first higgsino sign compared to the second one:\[
P=T_{3}=\left(\begin{array}{cc}
0 & 0\\
0 & -\sigma_{3}\end{array}\right)=\left(\begin{array}{cccc}
0 & 0 & 0 & 0\\
0 & 0 & 0 & 0\\
0 & 0 & -1 & 0\\
0 & 0 & 0 & 1\end{array}\right)\]
 we recover the gauge independence relation (\ref{general-gauge-relation}).
The particular form of the right hand side of this relation then follows
from the special structure of the symmetric neutralino mass matrix:
\begin{eqnarray*}
M & = & \left(\begin{array}{cc}
A & C\\
C^{T} & B\end{array}\right)\end{eqnarray*}
where $A$ is a $2\times2$ diagonal matrix reflecting the Majorana
nature of gauginos, $B$ is a $2\times2$ anti-diagonal matrix reflecting
the Dirac nature of the $SU(2)$-charged higgsino pair, and $C\sim m_{Z}$
is a $2\times2$ matrix that vanishes when $SU(2)$ is unbroken and
with null determinant to otherwise ensure masslessness of the photon.
These conditions make $PM+MP$ off-diagonal, explaining the non-trivial
vanishing of the left-hand side of (\ref{relamatrix}) with $m_{Z}$. 

These remarks stress the central role played by gauge invariance in
the structure of the mass matrix. The appearance of $P=T_{3}$ is
no surprise, as all $\chi\chi Z$ couplings find their root in the
gauge invariant $\tilde{h}\tilde{h}Z$: higgsinos can couple to the
$Z$ boson thanks to their $U_{Y}(1)$ charge which dictates a Dirac
behaviour. Only the spontaneous breaking of $SU(2)$ and $U_{Y}(1)$
carried by $C$ can then split this degenerate Dirac system into a
pair of Majorana particles.

\section{High Energy Unitarity}

Despite vague similarities, the gauge independence relations (\ref{general-gauge-relation})
do not yet explain the cancellation of (\ref{amplZ}) and (\ref{amplchi}),
and in particular their different powers of mixings $N_{ij}$. To
get further, we need another relation. Using the pinch technique\cite{Pinch},
we therefore turn to the high energy behaviour of the amplitude. Indeed,
from rotation invariance, the outgoing $Z$ boson must have a pure
longitudinal polarization, and it is well known that this can lead
to conflicts with the perturbative unitarity constraint that scattering
amplitudes be bounded by a constant $A(s\to\infty)<K$ . 

Having checked gauge independence, we can for simplicity use the Feynman
gauge $\xi=1$, to have a purely transverse $Z$ propagator free from
high energy divergences. The dangerous diagrams which must cancel
are then the $s$-channel Goldstone and $t$-channel neutralinos exchanges.

A light-like vector being orthogonal to itself, the polarization vector
at high energies is approximately\begin{eqnarray}
\varepsilon_{\mu}(k) & \simeq & \frac{k_{\mu}}{m_{Z}}.\label{HEapprox}\end{eqnarray}
Using this and the kinematic identity $2q.k=q^{2}+m_{Z}^{2}-m_{h}^{2}$,
the amplitude for Goldstone exchange (\ref{AmpGxi}) in this gauge
becomes:

\begin{eqnarray}
\begin{array}{c}
A\left(\chi\chi\rightarrow Zh\right)_{G}\end{array} & \begin{array}{c}
=\end{array} & \frac{ig^{2}}{2\cos^{2}\theta_{W}}\frac{O_{11}^{G}}{2m_{Z}}\ \overline{\chi}\gamma_{5}\chi\label{goldstone-pinch}\\
 &  & \hspace{2mm}\times\left(1+\frac{2m_{Z}^{2}-m_{h}^{2}}{q^{2}-m_{Z}^{2}}\right)\nonumber \end{eqnarray}

The first {}``contact'' term in the parenthesis gives a contribution
$A\simeq\sqrt{s}=\sqrt{q^{2}}$, which is divergent and violates unitarity
in the high energy limit, while the second term is better behaved
thanks to the appearance of a propagator denominator. From a diagrammatic
point of view, this is expressed by splitting the diagram into a {}``pinched''
part and a rest:

\begin{center}\includegraphics{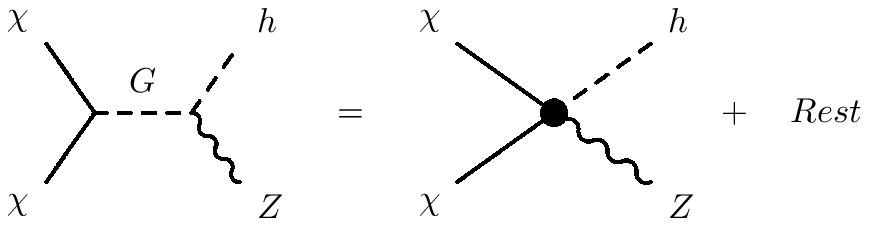}\par\end{center}

For the neutralino exchange channel, we have\begin{eqnarray*}
A\left(\chi\chi\rightarrow Zh\right)_{\chi} & = & -2i\sum_{i=1}^{4}\frac{C_{1i}^{A}C_{1i}^{h}}{Q^{2}-m_{\chi_{i}}^{2}}\\
 &  & \times\overline{\chi}\left(p'\right)\left(\slam{Q}+m_{\chi_{i}}\right)\gamma^{\mu}\gamma_{5}\chi\left(p\right)\ \frac{k_{\mu}}{m_{Z}}\end{eqnarray*}
Expressing $\slam{k}=\slam{p}-\slam{Q}=(\slam{p}-m_{\chi})-(\slam{Q}-m_{\chi_{i}})+(m_{\chi}-m_{\chi_{i}})$,
the first term vanishes on-shell, while the second cancels the propagator
pole to give a contact term and a rest:

\begin{eqnarray}
A\left(Zh\right)_{\chi} & = & \frac{-ig^{2}}{2c_{W}^{2}}\sum_{i=1}^{4}\frac{O_{1i}^{Z}O_{1i}^{h}}{m_{Z}}\times\{\overline{\chi}\gamma_{5}\chi\label{neutralino-pinch}\\
 &  & \hphantom{-i2}-\frac{m_{\chi}-m_{\chi_{i}}}{Q^{2}-m_{\chi_{i}}^{2}}\ \overline{\chi}\left(\slam{Q}+m_{\chi_{i}}\right)\gamma_{5}\chi\}\nonumber \end{eqnarray}
which can be diagrammatically represented by:

\begin{center}\includegraphics{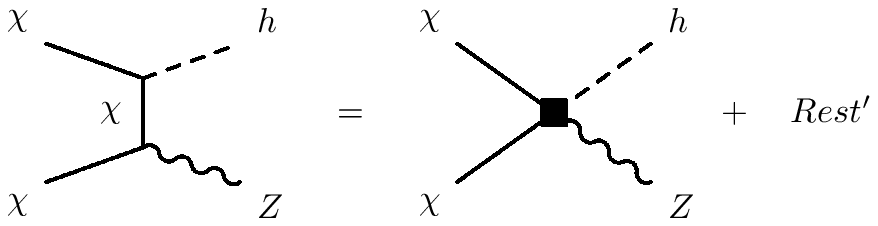}\par\end{center}

Cancellation of the contact terms in (\ref{goldstone-pinch}) and
(\ref{neutralino-pinch}), requires the following identity: \begin{eqnarray*}
\frac{1}{2}O_{11}^{G} & = & \sum_{i}O_{1i}^{Z}O_{1i}^{h}\end{eqnarray*}
 or, using (\ref{OZij}), (\ref{Ohij}) and (\ref{OGij}), the following
relation among mixings $N$:\begin{eqnarray}
 &  & \begin{array}{c}
\left(N_{12}c_{W}-N_{11}s_{W}\right)\left(s_{\beta}N_{14}+c_{\beta}N_{13}\right)\end{array}\label{rela-pinch}\\
 &  & \hspace{1em}=\sum_{i=1}^{4}\left(N_{14}N_{i4}-N_{13}N_{i3}\right)\times\nonumber \\
 & \  & \hphantom{\hspace{1em}=\sum_{i=1}(}\{\left(c_{W}N_{12}-s_{W}N_{11}\right)\left(N_{i4}s_{\beta}-N_{i3}c_{\beta}\right)\nonumber \\
 & \  & \hphantom{\hspace{1em}=\sum_{i=1}(\{(}+\left(c_{W}N_{i2}-s_{W}N_{i1}\right)\left(N_{14}s_{\beta}-N_{13}c_{\beta}\right)\}\nonumber \end{eqnarray}
As complicated as it may seen, this equation simply follows from the
orthogonality condition $N_{ij}N_{kj}=\delta_{ik}$, with $i,k=3,4$.
We have therefore shown that high energy perturbative unitarity of
the amplitude is guaranteed by the unitarity of the mixing matrix.

By combining (\ref{rela-pinch}) and (\ref{rela-gauge}), a new non-trivial
identity among couplings is obtained: \begin{eqnarray*}
\frac{m_{\chi}}{m_{Z}}O_{11}^{Z} & = & -\sum_{i}O_{1i}^{Z}O_{1i}^{h}\end{eqnarray*}
which is even less trivial when using the definitions of $O_{ij}^{Z}$
and $O_{ij}^{h}$:

\begin{eqnarray}
\frac{m_{\chi}}{m_{Z}}\left(N_{i4}^{2}-N_{i3}^{2}\right) & = & \begin{array}{c}
-\sum_{i=1}^{4}\left(N_{14}N_{i4}-N_{13}N_{i3}\right)\end{array}\label{relationcoupling}\\
 &  & \times\{\left(c_{W}N_{12}-s_{W}N_{11}\right)\left(N_{i4}s_{\beta}-N_{i3}c_{\beta}\right)\nonumber \\
 &  & +\left(c_{W}N_{i2}-s_{W}N_{i1}\right)\left(N_{14}s_{\beta}-N_{13}c_{\beta}\right)\}\nonumber \end{eqnarray}
This identity relates the couplings appearing in $Z$-exchange (\ref{amplZ})
and $\chi$-exchange (\ref{amplchi}), and suggests to rewrite the
second as:

\begin{eqnarray}
A\left(\chi\chi\rightarrow Zh\right)_{\chi} & = & 2i\sqrt{2}\beta_{Zh}\frac{m_{\chi}}{m_{Z}}\frac{g^{2}}{cos^{2}\theta_{W}}\nonumber \\
 &  & \times\left(\frac{m_{\chi}}{m_{Z}}O_{11}^{Z}+\sum_{i=1}^{4}R_{i}\right)\label{1+Ri}\end{eqnarray}
where the first term exactly cancels the $s$-channel $Z$ exchange,
and the second is subdominant in the high energy limit $s\gg m_{Z}^{2},m_{\chi}^{2}$.
It is however not clear why those remainders should be subdominant
at rest, because even for large $s=4m_{\chi}^{2}$, there is confusion
between dynamically suppressed contributions $\sim m_{Z}^{2}/s\ll1$
and neutralino mixing suppressed ones $\sim m_{Z}^{2}/m_{\chi}^{2}\ll1$.
Moreover, the definition

\begin{eqnarray}
R_{i} & = & O_{1i}^{Z}O_{1i}^{h}\frac{\left(2m_{\chi_{i}}m_{\chi}-4m_{\chi}^{2}-2m_{\chi_{i}}^{2}+m_{h}^{2}+m_{Z}^{2}\right)}{2m_{\chi}^{2}+2m_{\chi_{i}}^{2}-m_{h}^{2}-m_{Z}^{2}}\quad\label{Ri}\end{eqnarray}
implies that for $m_{Z}\ll m_{\chi}$, $R_{3}\approx-R_{4}$ are comparable
with the first term in (\ref{1+Ri}), and only their sum becomes negligible.
To analyze this final cancellation, a perturbative expansion in $m_{Z}/m_{\chi}$
is therefore needed.

\section{Perturbation theory}

In the SUSY-decoupling limit ($m_{Z}\ll M_{1},M_{2},\mu$), the neutralino
mass in Appendix \ref{sec:Neutralino-Mass-Matrix} is naturally split\cite{gunion-haber}
into $M=M_{0}+W$, with a leading contribution

\begin{eqnarray*}
M_{0} & = & \left(\begin{array}{cccc}
M_{1} & 0 & 0 & 0\\
0 & M_{2} & 0 & 0\\
0 & 0 & 0 & -\mu\\
0 & 0 & -\mu & 0\end{array}\right)\end{eqnarray*}
and a perturbation

\begin{eqnarray*}
W & = & m_{Z}\left(\begin{array}{cccc}
0 & 0 & -s_{W}c_{\beta} & s_{W}s_{\beta}\\
0 & 0 & c_{W}c_{\beta} & -c_{W}s_{\beta}\\
-s_{W}c_{\beta} & c_{W}c_{\beta} & 0 & 0\\
s_{W}s_{\beta} & -c_{W}s_{\beta} & 0 & 0\end{array}\right)\end{eqnarray*}
triggered by EW-symmetry breaking. Following standard Rayleigh-Schroedinger
perturbation expansion we start by solving the unperturbed eigensystem\begin{eqnarray*}
N^{0}M_{0}N^{0\, T} & = & m^{0}\end{eqnarray*}
 whose eigenvalues are $m^{0}=diag\left(M_{1},M_{2},-\mu,\mu\right)$
and whose eigenvectors $\varphi_{n}^{0}$ form the mixing matrix\[
N^{0\, T}=(\varphi_{1}^{0},\varphi_{2}^{0},\varphi_{3}^{0},\varphi_{4}^{0})=\left(\begin{array}{cccc}
1 & 0 & 0 & 0\\
0 & 1 & 0 & 0\\
0 & 0 & \frac{1}{\sqrt{2}} & \frac{-1}{\sqrt{2}}\\
0 & 0 & \frac{1}{\sqrt{2}} & \frac{1}{\sqrt{2}}\end{array}\right)\]
The first order corrections to eigenvalues $m_{n}^{1}=\left<\varphi_{n}^{0}\right|W\left|\varphi_{n}^{0}\right>$
are simply the diagonal elements of $W^{0}=N^{0}W\ N^{0\, T}$:

\begin{eqnarray}
W^{0} & = & m_{Z}\left(\begin{array}{cccc}
0 & 0 & \frac{s_{W}s_{-}}{\sqrt{2}} & \frac{s_{W}s_{+}}{\sqrt{2}}\\
0 & 0 & -\frac{c_{W}s_{-}}{\sqrt{2}} & -\frac{c_{W}s_{+}}{\sqrt{2}}\\
\frac{s_{W}s_{-}}{\sqrt{2}} & -\frac{c_{W}s_{-}}{\sqrt{2}} & 0 & 0\\
\frac{s_{W}s_{+}}{\sqrt{2}} & -\frac{c_{W}s_{+}}{\sqrt{2}} & 0 & 0\end{array}\right)\label{eigenvlues1order}\end{eqnarray}
where $s_{\pm}=s_{\beta}\pm c_{\beta}$. From the structure of the
perturbation $W$, these diagonal elements clearly vanish. Furthermore,
first order corrections to eigenvectors

\begin{eqnarray*}
\left|\varphi_{i}^{1}\right> & = & \sum_{j\neq i}\frac{W_{ji}^{0}}{m_{i}^{0}-m_{j}^{0}}\left|\varphi_{j}^{0}\right>\end{eqnarray*}
can be regrouped into $N^{1\, T}=(\varphi_{1}^{1},\varphi_{2}^{1},\varphi_{3}^{1},\varphi_{4}^{1})$
with:

\[
N^{1}=m_{Z}\left(\begin{array}{cccc}
0 & 0 & -\frac{s_{W}C_{1}}{M_{1}^{2}-\mu^{2}} & \frac{s_{W}S_{1}}{M_{1}^{2}-\mu^{2}}\\
0 & 0 & \frac{c_{W}C_{2}}{M_{2}^{2}-\mu^{2}} & -\frac{c_{W}S_{2}}{M_{2}^{2}-\mu^{2}}\\
-\frac{s_{W}s_{-}}{\sqrt{2}\left(M_{1}+\mu\right)} & \frac{c_{W}s_{-}}{\sqrt{2}\left(M_{2}+\mu\right)} & 0 & 0\\
\frac{s_{W}s_{+}}{\sqrt{2}\left(\mu-M_{1}\right)} & \frac{c_{W}s_{+}}{\sqrt{2}\left(M_{2}-\mu\right)} & 0 & 0\end{array}\right)\]
and $C_{1,2}=\left(\mu s_{\beta}+M_{1,2}c_{\beta}\right)$, $S_{1,2}=\left(M_{1,2}s_{\beta}+\mu c_{\beta}\right)$.

For a bino-like neutralino ($M_{1}<M_{2},\mu)$, the $Z\chi\chi$
vertex does not exist without electroweak symmetry breaking and knowing
the diagonalization matrix $N=N^{0}+N^{1}$ up to first order in $m_{Z}$
allows in fact to compute the first non-trivial contribution to $Z$-exchange
(\ref{amplZ}) which appears at second order:\begin{eqnarray}
A\left(\chi\chi\rightarrow Zh\right)_{Z} & = & -\sqrt{2}\beta_{Zh}\frac{M_{1}^{2}}{m_{Z}^{2}}g^{2}c_{2\beta}t_{W}^{2}\frac{m_{Z}^{2}}{\left(M_{1}^{2}-\mu^{2}\right)}\label{perturbedZ}\end{eqnarray}
At the same order, the first non-trivial contribution to $\chi$-exchange
(\ref{amplchi}) requires only one perturbation of the $Z\chi\chi_{i}$
vertex, and no perturbation of the $h\chi\chi_{i}$ vertex, which
does exist in the absence of electroweak symmetry breaking. Further
expanding the propagator in powers of $m_{Z}$ would give the same
structure as (\ref{1+Ri}):\begin{eqnarray}
A\left(\chi\rightarrow Zh\right)_{\chi} & = & \sqrt{2}\beta_{Zh}\frac{M_{1}^{2}}{m_{Z}^{2}}g^{2}\cos2\beta\tan^{2}\theta_{W}\label{perturbedchiexpansion}\\
 &  & \times\frac{m_{Z}^{2}}{\left(M_{1}^{2}-\mu^{2}\right)}\left(1+\frac{1}{2}\frac{m_{Z}^{2}+m_{h}^{2}}{M_{1}^{2}+\mu^{2}}\right),\nonumber \end{eqnarray}
suggesting that the remainder is $O(m_{Z}^{4})$, \emph{i.e.} two
orders lower than the leading term. However, to firmly establish this
conclusion requires a justification for the absence of terms $O(m_{Z}^{3})$,
which we shall now find by examining the general structure of the
perturbative expansion.

When solving the eigensystem \begin{eqnarray}
\left(M_{o}+W\right)_{ij}N_{jl}^{T} & = & N_{il}^{T}m_{l}\label{eigensystem}\end{eqnarray}
by power expansions in $m_{Z}$ for eigenvalues and eigenvectors:\begin{eqnarray}
m_{i} & = & m_{i}^{0}+m_{i}^{1}+...\label{mexpansion}\\
N & = & N^{0}+N^{1}+...\label{Nexpansion}\end{eqnarray}
we can to all orders choose the correction to an eigenvector in $N-N^{0}$
orthogonal to the corresponding unperturbed vector in $N^{0}$: the
only price is to end up with non-unit vectors in $N$. This choice
however simplifies the recurrence relation for the solution at order
$q$ to: 

\begin{eqnarray}
m_{i}^{q} & = & \left(N^{0}WN^{q-1\, T}\right)_{ii}\label{mq}\\
\left(N^{0}N^{q\, T}\right)_{ji} & = & \frac{\left(N^{0}WN^{q-1\, T}\right)_{ji}}{m_{i}^{0}-m_{j}^{0}}-\sum_{p=1}^{q-1}m_{i}^{p}\frac{\left(N^{0}N^{q-p\, T}\right)_{ji}}{m_{i}^{0}-m_{j}^{0}}\nonumber \end{eqnarray}
The expressions for $q=0$, and $q=1$ were given above. We saw that
$M_{0}$ is $2\times2$ block-diagonal, and so is $N^{0}$, whereas
$W$ and thus $N^{1}$ are block off-diagonal. Following the recurrence,
this can be generalized, to show that $N^{q}$ must be block diagonal
for $q$ even and block off-diagonal for $q$ odd. Because of this
structure, $m_{i}^{q}$ will vanish for $q$ odd, so that $m_{i}(m_{Z})$
is holomorphic in $m_{Z}^{2}$. In a similar way, diagonal blocks
of $N$ have a purely even power expansion in $m_{Z}$, whereas off-diagonal
ones only contain odd powers.

These results can be extended to the amplitudes $A_{Z}$ and $A_{\chi}$
which then contain only even powers of $m_{Z}$: the lowest order
$O(m_{Z}^{-2})$ vanishes for both as it should to allow for a smooth
$m_{Z}\to0$ limit, the next $O(m_{Z}^{0})$ is equal and opposite
for $s$- and $t$-channel exchanges, and the remainder $O(m_{Z}^{2})$
dictates the amplitude of the $Zh$ annihilation channel at rest to
be lower than the $t\bar{t}$ pair channel. This can be loosely expressed
as:

\[
A\left(Zh\right)\propto\frac{m_{Z}^{2}}{m_{\chi}^{2}}\ll A\left(t\overline{t}\right)\propto\frac{m_{t}}{m_{\chi}}\ll A\left(Zh\right)_{Z}\propto1\]
The order of magnitude and the power of the suppression\begin{eqnarray*}
\frac{\sigma\left(\chi\chi\rightarrow Zh\right)_{Z+\chi}}{\sigma\left(\chi\chi\rightarrow Zh\right)_{Z}} & \propto & \frac{m_{Z}^{4}}{M_{1}^{4}}\end{eqnarray*}
 then agree with those displayed in figs.~\ref{Zsursigamtot},\ref{MSSMscan}.

\section{Conclusion}

In this work, we have given a quantitative understanding of why neutralinos
at rest cannot annihilate predominantly into $Zh$. The possible relevance
of this process for indirect DM detection has been shown in the Introduction,
before pointing out similarities and differences between the $t\overline{t}$
and $Zh$ annihilation channels. We also stressed how much a naive
estimate of this last channel can fail by ignoring the subtle but
tight links between couplings imposed by symmetries, especially broken
ones. Because of these links, a fine cancellation does occur which
requires a closer look. Having noticed that this cancellation was
getting finer with increasingly broken supersymmetry, we showed that
broken gauge symmetry had to be investigated. This was done both at
the level of gauge independence in $R_{\xi}$ gauges, and of unitarity
at high energy, known to be delicate for longitudinal gauge bosons.
Both constraints led to non-trivial relations among couplings which
showed that indeed, the SM particles exchanges can be cancelled by
superpartners exchanges, as heavy as these might be. However, to quantitatively
estimate the importance of what remains after this cancellation required
to show that a perturbative expansion of the amplitudes in $m_{Z}/m_{\chi}$
contained only even powers. Whether such cancellation can be extended
from large $s$ to large $m_{\chi}=\sqrt{s}/2$ at rest for all diagrams
suffering from large $s$ unitarity problems, remains an open question.

\begin{acknowledgement}
It is a pleasure to acknowledge fruitful discussions with A. Djouadi,
P. Gay, R. Grimm, J.-L. Kneur, Y. Mambrini, V. Morenas, G. Moultaka
and J. Papavassiliou. This work was partially funded by the GdR 2305
{}``Supersymétrie'' of the French CNRS.
\end{acknowledgement}
\appendix

\section{Indirect Detection Energy Spectra\label{sec:Indirect-Detection-Energy}}

The neutrino and photon differential energy spectra of figure~1 were
extracted from a PYTHIA simulation of $10^{6}$ events for each channel,
shown as a function of $x=E_{\nu,\ \gamma}/m_{\chi}$. For the hard
(anti-)neutrinos from the $Z$-boson in the $Zh$ channel that concerned
us most, we have been careful to correct the unpolarized PYTHIA results
by a factor $\propto x(1-x)$ translating the purely longitudinal
polarization of the $Z$-boson, which suppresses forward neutrinos
with respect to the $WW$ channel. In spite of this factor, the $Zh$
channel produces the next-to-hardest neutrino spectrum.

For photons, the hard (flat) component of the $Zh$ channel around
$x\approx1$ comes from $h$ loop-decaying into two photons and therefore
only appears at the largest values of $x$. This leaves only a small
number of events (\textasciitilde{}10/bin) and large statistical errors
which do not appear in the fit shown in the bottom figure 1. There
are of course no events for $x>1$, but the precise shape of this
vanishing (probably similar to that of neutrinos from $WW$ above,
as shown) is hidden by these errors.

\section{\label{sub:Dirac-Majorana}Dirac and Majorana fermion conventions}

To represent the Clifford algebra of Dirac matrices \begin{eqnarray*}
\left\{ \gamma^{\mu},\gamma^{\nu}\right\}  & = & 2\eta^{\mu\nu}\end{eqnarray*}
we use the chiral basis: 

\[
\begin{array}{ccc}
\gamma^{\mu}=\left(\begin{array}{cc}
0 & \sigma^{\mu}\\
\overline{\sigma}^{\mu} & 0\end{array}\right) &  & \gamma_{5}=\left(\begin{array}{cc}
-1 & 0\\
0 & 1\end{array}\right)\end{array}\]
where

\begin{eqnarray*}
\sigma^{\mu}=\left(1,\vec{\sigma}\right) &  & \overline{\sigma}^{\mu}=\left(1,-\vec{\sigma}\right)\end{eqnarray*}
are 4-d extensions of the Pauli matrices $\vec{\sigma}=\left(\sigma^{1},\sigma^{2},\sigma^{3}\right)$.

For Majorana spinors, we use\cite{Haber} the usual Dirac Feynman
rules after having chosen an arbitrary orientation of the spinor lines.
The Majorana condition is\begin{eqnarray}
\chi= & \chi^{c}= & C\overline{\chi}^{T}\label{majorana-condition}\end{eqnarray}
and the plane wave expansion of a Majorana field operator is thus
\cite{mohapatra}: \begin{eqnarray}
\chi\left(x\right) & = & \int\frac{d^{3}p}{\left(2\pi\right)^{3/2}}\sum_{s=\pm}\left(a_{s}\left(p\right)u_{s}\left(p\right)e^{-ip.x}\right.\label{majorana-field}\\
 &  & \hphantom{\int\frac{d^{3}p}{\left(2\pi\right)^{3/2}}\sum_{s=\pm}}\left.+a_{s}^{\dagger}\left(p\right)v_{s}\left(p\right)e^{ip.x}\right)\nonumber \end{eqnarray}
Implementing the Majorana condition (\ref{majorana-condition}) in
(\ref{majorana-field}), we find the relations\begin{eqnarray}
u_{s} & = & C\gamma^{0}v_{s}^{*}\label{majorana-u}\\
v_{s} & = & C\gamma^{0}u_{s}^{*}\label{majorana-v}\end{eqnarray}
which allow to flip the orientation of external Majorana lines.

\section{Neutralino Mass Matrix\label{sec:Neutralino-Mass-Matrix}}

The neutralino mass eigenstates are linear combinations of gaugino
and higgsino fields $\left(\tilde{B},\tilde{W}_{3},\tilde{H}_{b},\tilde{H}_{t}\right)$

\begin{eqnarray*}
\chi_{i} & = & N_{i1}\tilde{B}+N_{i2}\tilde{W}_{3}+N_{i3}\tilde{H}_{b}+N_{i4}\tilde{H}_{t}\end{eqnarray*}
which diagonalize the mass matrix: 

\begin{eqnarray}
M & = & \left(\begin{array}{cc}
\begin{array}{cc}
M_{1} & 0\\
0 & M_{2}\end{array} & m_{Z}\times C\\
m_{Z}\times C^{T} & \begin{array}{cc}
0 & -\mu\\
-\mu & 0\end{array}\end{array}\right)\label{massmatrix}\end{eqnarray}
$C$ is the $2\times2$ electroweak-breaking contribution to this
neutralino mass matrix :

\begin{eqnarray*}
C & = & \left(\begin{array}{cc}
-s_{W}c_{\beta} & s_{W}s_{\beta}\\
c_{W}c_{\beta} & -c_{W}s_{\beta}\end{array}\right)\end{eqnarray*}

with

\begin{center}$\begin{array}{cc}
s_{W}=\sin\theta_{W}, & c_{W}=\cos\theta_{W}\\
s_{\beta}=\sin\beta, & c_{\beta}=\cos\beta\end{array}$ \par\end{center}

The normalized eigenvectors can be collected into a unitary matrix
$N$ satisfying:

\begin{eqnarray}
N^{\star}MN^{-1} & = & M_{D}\label{diagonalization-formula}\end{eqnarray}
where $M_{D}$ is a diagonal matrix containing neutralino masses.

In the absence of CP violating phases, the matrix $N$ can be chosen
as a real matrix, and at least one neutralino mass is then negative.

\section{Lagrangian terms for relevant couplings\label{sec:Lagrangian-terms-for}}

\subsection{$Z-\chi-\chi$ coupling\label{sub:Z-chi-chi-coupling}}

\begin{eqnarray*}
\mathcal{L} & = & \frac{1}{2}\sum_{i,j=1}^{4}\overline{\chi_{i}}\gamma^{\mu}\left(C_{ij}^{V}-C_{ij}^{A}\gamma_{5}\right)\chi_{j}Z_{\mu}\end{eqnarray*}
with \begin{eqnarray}
C_{ij}^{V} & = & \frac{g}{4\cos\theta_{W}}\left(O_{ij}^{Z}-O_{ij}^{Z*}\right)\label{Cv}\\
C_{ij}^{A} & = & \frac{g}{4\cos\theta_{W}}\left(O_{ij}^{Z}+O_{ij}^{^{Z}*}\right)\label{Ca}\end{eqnarray}
and 

\begin{eqnarray*}
O_{ij}^{Z} & = & N_{i4}N_{j4}^{*}-N_{i3}N_{j3}^{*}\end{eqnarray*}
In the absence of CP violation, there is a basis such that $C_{ij}^{V}=0$.

\subsection{$h$-$Z$-$Z$ coupling}

\begin{eqnarray}
\mathcal{L} & = & \frac{1}{2}C_{Z}hZ_{\mu}Z^{\mu}\label{sin(alpha-beta)}\end{eqnarray}
where 

\begin{eqnarray}
C_{Z} & \ = & -\frac{gm_{Z}}{\cos\theta_{W}}\sin\left(\alpha-\beta\right)\label{Cz}\end{eqnarray}
In practice, we will always work in the decoupling limit $m_{A}\gg m_{Z}$,
which implies: \begin{eqnarray}
\alpha & = & \beta-\frac{\pi}{2}\label{decoupling-condition}\end{eqnarray}
Hence, we have $\sin\left(\alpha-\beta\right)\simeq-1$ in (\ref{Cz}).

\subsection{$h-\chi-\chi$ coupling}

\begin{eqnarray}
\mathcal{L} & = & \frac{g}{2}h\overline{\chi}_{i}\left\{ s_{\alpha}\left(L_{ij}^{*}P_{L}+L_{ij}P_{R}\right)\right.\label{L h-chi-chi}\\
 &  & \hphantom{\frac{g}{2}h\overline{\chi}_{i}}\left.+c_{\alpha}\left(K_{ij}^{*}P_{L}+K_{ij}P_{R}\right)\right\} \chi_{j}\nonumber \end{eqnarray}

with \begin{eqnarray*}
K_{ij} & = & \frac{1}{2}N_{i4}\left(N_{j2}-\tan\theta_{W}N_{j1}\right)+\left(i\leftrightarrow j\right)\\
L_{ij} & = & \frac{1}{2}N_{i3}\left(N_{j2}-\tan\theta_{W}N_{j1}\right)+\left(i\leftrightarrow j\right)\end{eqnarray*}

The symmetric form of $K$ and $L$ comes from the fact we are working
with Majorana particles: 

\begin{eqnarray*}
\overline{\chi}_{i}\left(1\pm\gamma_{5}\right)\chi_{j} & = & \overline{\chi}_{j}\left(1\pm\gamma_{5}\right)\chi_{i}\end{eqnarray*}

For real $N$, this Lagrangian simplifies to:

\begin{eqnarray*}
\mathcal{L} & = & \frac{1}{2}\sum_{i,j=1}^{4}C_{ij}^{h}h\overline{\chi}_{i}\chi_{j}\end{eqnarray*}

where\begin{eqnarray}
C_{ij}^{h} & = & \frac{g}{2\cos\theta_{W}}O_{ij}^{h}\label{Ch}\end{eqnarray}

and \begin{eqnarray*}
O_{ij}^{h} & = & \left(c_{W}N_{i2}-s_{W}N_{i1}\right)\left(s_{\alpha}N_{j3}+c_{\alpha}N_{j4}\right)+\left(i\leftrightarrow j\right)\end{eqnarray*}

\subsection{$Z$-fermion-fermion coupling}

\begin{eqnarray*}
\mathcal{L} & = & \sum_{f}\overline{f}\gamma^{\mu}\left(C_{ff}^{ZV}-C_{ff}^{ZA}\gamma_{5}\right)fZ_{\mu}\end{eqnarray*}
where \begin{eqnarray*}
C_{ff}^{ZV} & = & -\frac{g}{2\cos\theta_{W}}\left(T_{3f}-2\sin^{2}\theta_{W}Q_{f}\right)\\
C_{ff}^{ZA} & = & -\frac{g}{2\cos\theta_{W}}T_{3f}\end{eqnarray*}
$Q_{f}$ and $T_{3f}$ are the charge and third weak isospin component,
with the usual normalization: $T_{3top}=1$, and $Q_{top}=\frac{2}{3}$.

\end{document}